\input{aipcheck}

\documentclass[
  ,final            
  ]
  {aipproc}

\layoutstyle{8x11double}

\usepackage{subfigure}

\begin{document}

\title{Hadronic Production of Gamma Rays and\\ Starburst Galaxies}


\author{Niklas Karlsson}{
  address={Adler Planetarium, Chicago, IL 60605, USA}
}

\begin{abstract}
The Milky Way has been estabished to emit gamma rays. These gamma rays are presumably dominated by decays of neutral pions, although inverse Compton scatterings and bremsstrahlung also contribute. It is plausible that other galaxies can be diffuse sources of gamma rays in a similar manner. Starburst galaxies are particularly interesting to study as they are expected to have much higher cosmic-ray fluxes and interstellar matter densities. The neutral pions are created in cosmic-ray interactions with interstellar matter. Presented here is an overview of the recent work by Karlsson and co-workers on proton-proton interactions and the resulting secondary particle inclusive cross sections and angular distributions. This model can be used to calculated the $\pi^{0}$ component of the gamma-ray yield and spectrum from a starburst galaxy. The yield is expected to increase significantly (30\% to 50\%) and the spectrum to be harder than the incident proton spectrum. 

\end{abstract}

\maketitle


\section{Introduction}
The Milky Way Galaxy is a known source of gamma rays in MeV to GeV energies. This emission is particularly strong in the Galactic disk and traces closely the distribution of gas in the Galaxy. It is presumably dominated by gamma rays from decaying neutral pions, created in interactions between cosmic rays and interstellar matter (see e.g. \citep{article:Strong_etal:2000}). The gamma-ray flux and spectral shape measured by the Energetic Gamma-Ray Experiment Telescope (EGRET) \citep{article:Hunter_etal:1997} is considered the key attestation of this interpretation. Contributions also come from inverse Compton scatterings and bremsstrahlung due to electrons.

It is plausible that other galaxies with properties similar to the Milky Way also emit gamma rays, but the extremely large distances involved and low expected luminosities make detections very unlikely. To date, the Large Magellanic Cloud is the only external galaxy detected as a diffuse source of gamma rays \citep{article:Sreekumar_etal:1992}.

In a starburst galaxy there are regions with star-formation rates several orders of magnitude higher than that of normal galaxies. With an enhanced star-formation rate follows an enhanced rate of supernovae explosions and the cosmic-ray flux is expected to be higher (under the assumption of cosmic-ray acceleration in supernova remnants). Considering a galaxy with much higher cosmic-ray flux and a denser interstellar medium, compared to the Milky Way, the gamma-ray luminosity may be high enough to allow for a detection, even in TeV gamma rays, despite the distances involved.

Presented here is the parameterization of gamma-ray inclusive cross sections \citep{article:Kamae_etal:2006} and angular distributions \citep{article:KarlssonKamae:2008} produced by proton-proton interactions in astronomical environments. The parameterization is derived from simulations using the Pythia Monte Carlo event generator. The parametric model can be used to model the hadronic component of the gamma-ray emission from a typical starburst galaxy and allows accurate calculations of the expected gamma-ray flux and spectral shape. The fluctuations in the histograms are due to low event statistics.

\section{Hadronic Production of Gamma Rays}
Pions are produced in hadronic interactions between cosmic rays and interstellar matter. The major contribution comes from proton-proton ($p$-$p$) interactions, but there are also contributions from heavier nuclei. The neutral pion, $\pi^{0}$, decays predominantly into two gamma rays and the charged pions decay into muons, which subsequently decay into electrons/positrons, and neutrinos.

\begin{figure}
\label{fig:sigma_pp}
\includegraphics[width=3.1in]{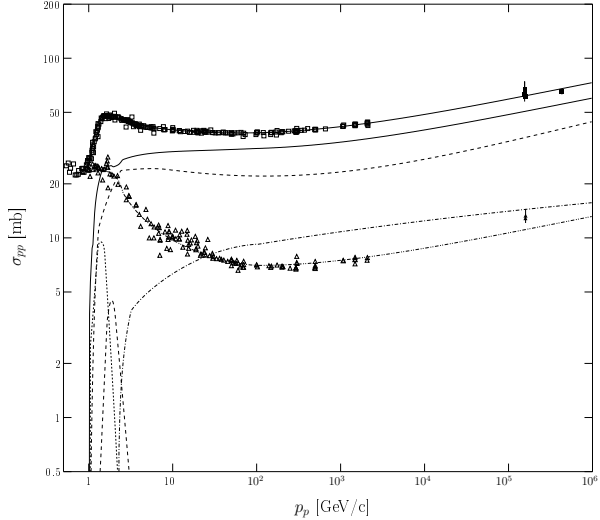}
\caption{Experimental $p$-$p$ ($\bar{p}$-$p$) cross sections, total (squares) and elastic (triangles), as functions of proton laboratory momentum, and their model fits; total (thin solid) and elastic (dot-dot dashed). The other lines are: total inelastic (thick solid), non-diffractive (dashed), diffractive (dot-dashed), $\Delta(1232)$ (dotted), and res(1600) (thin dashed).}
\end{figure}

The work by \citet{article:KamaeAbeKoi:2005} showed that previous models (see references in \citep{article:Kamae_etal:2006, article:KarlssonKamae:2008}) of pion production left out two important features of the inelastic $p$-$p$ interaction; the diffraction dissociation process and the Feynman scaling violation. It was also noted that the inelastic $p$-$p$ cross section was assumed to be constant, about 24 mb, for $T_{p}\gg 10$ GeV, in contradiction to recent experimental data, where a logarithmic increase with proton energy is evident \citep{article:Hagiwara_etal:2002}. This is shown in Figure \ref{fig:sigma_pp} together with the component cross sections used in this work.

The model by \citet{article:KamaeAbeKoi:2005} included only the non-diffractive interaction and the diffraction dissociation process. This model is not accurate near the pion production threshold (see Figure 5 in \citet{article:KamaeAbeKoi:2005}). To correct this and improve accuracy the contribution of two baryon resonance excitation states, the physical $\Delta(1232)$ resonance and res(1600) representing several resonances around 1600 MeV/c$^{2}$, were included. Figure \ref{fig:sigma_pi0} shows the improved total inclusive $\pi^{0}$ cross section. Further necessary adjustments to model A required by the baryon resonances are described in section 3 of \citet{article:Kamae_etal:2006}.

The interaction model presented here does not include $\alpha$-$p$, $p-$He, and $\alpha$-He due to paucity of experimental data. To an approximation, the $\alpha$-particle and the He nucleus can be taken as four individual nucleons. The error is expected to be less than 10\% for high-energy gamma rays.

\subsection{Monte Carlo Simulations}
The parameterizations of inclusive cross sections and angular distributions are derived from Monte Carlo simulations of the adjusted $p$-$p$ interaction model described in \citet{article:Kamae_etal:2006}. Events were generated for each of the four components at discrete proton kinetic energies (0.488 GeV $\leq T_{p}\leq$ 512 TeV) from a geometrical series
\begin{equation}
T_{p}=1000\times 2^{(i-22)/2}\ \mathrm{GeV},\ i=0,\ldots,40. 
\end{equation}
Each proton kinetic energy, $T_{p}$, represents a bin covering $2^{-0.25}T_{p}$ to $2^{0.25}T_{p}$. The addition of the resonances to the model required an increased sampling frequency near the pion production threshold and events were also generated for $T_{p}=0.58$ and 0.82 GeV. Events were not generated for proton energies where the component cross section is very small or zero.

\begin{figure}
\label{fig:sigma_pi0}
\includegraphics[width=3in]{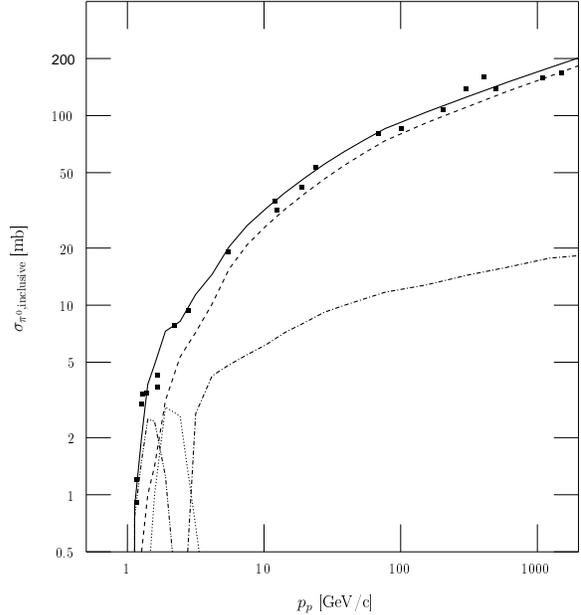}
\caption{Experimental (squares) and Monte Carlo simulated (solid) total inclusice $\pi^{0}$ cross section as functions of proton laboratory momentum. The other lines are for the four components of the $p$-$p$ interaction model: non-diffractive (dashed), diffractive process (dot-dashed), $\Delta(1232)$ (double-dot dashed), and res(1600) (dotted).}
\end{figure}

The results of these Monte Carlo simulations have been verified to agree with experimental data available for pions. The total inclusive $\pi^{0}$ cross section agrees with the experimental one (see Figure \ref{fig:sigma_pi0}) and  the simulations were ensured to reproduce the distributions of pion kinetic energy in the $p$-$p$ center-of-mass (CM) system in the resonance region (see Figures 3,4 and 5 in \citep{article:Kamae_etal:2006}). The angular distribution was verified using experimental data on the Lorentz invariant cross section, $Ed^{3}\sigma/dp^{3}$, and the average transverse momentum of pions (see Figures 2 and 3 in \citep{article:KarlssonKamae:2008}). 

\subsection{Parameterizations}
The inclusive cross sections, $\Delta\sigma(E,T_{p})/\Delta\log{E}$, of all stable secondary particles, gamma rays, electrons, positrons, electron and muon neutrinos (and their anti-particles) have been parameterized. For each component, the mono-energetic cross section $\Delta\sigma(E)/\Delta\log{E}$ were fitted to a common algebraic form $F(x)$ with $x=\log{E}$ and $E$ in GeV. Each fit gives a set of parameters which were then fitted as functions of $T_{p}$ to give the final parameterization of the inclusive cross section. The algebraic forms and the fits are described in \citet{article:Kamae_etal:2006}.

\begin{figure}
\label{fig:gamma_spectrum}
\includegraphics[width=3.1in]{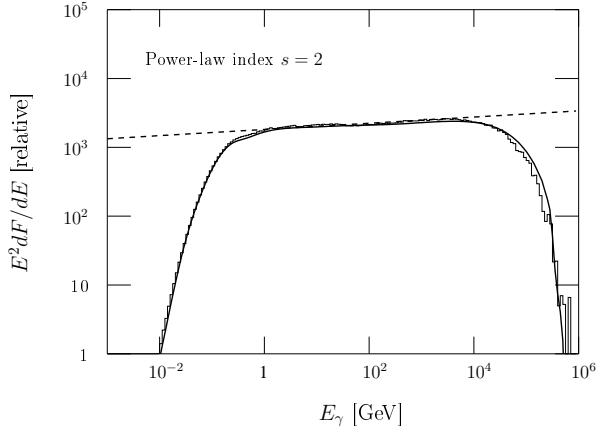}
\caption{Gamma-ray spectrum produced by protons with a power-law spectrum of index 2.0 and cutoff at $T_{p}=512$ TeV. The solid line was calculated using the parameterization described in the text and the histogram was derived from the Monte Carlo simulated events. The dashed line is an eyeball fit to a power law with index 1.95. The absolute normalization is relative to the density of target protons.}
\end{figure}

The gamma-ray spectrum, $E^{2}dF/dE$, produced by a power-law distribution, index 2.0, of protons was calculated using the parameterization and compared with the corresponding Monte Carlo spectrum in Figure \ref{fig:gamma_spectrum}. The parameterization is accurate to within 10\% except near the upper kinematic limit where the error is up to 20\%.

The angular distribution, i.e.\ the distribution over the polar angle $\theta$, is readily derived from the $p_{t}$ distribution. 
The latter is expected to follow a simple exponential form. Thus, for gamma rays, the $p_{t}$ distribution was fitted to an exponential for each $T_{p}$ and also the gamma-ray energy $E$. Again, each fit gives a set of parameters which were fitted as functions of $T_{p}$ and $E$ to give the final parameterization of the $p_{t}$ distribution. The details of this procedure are described in \citet{article:KarlssonKamae:2008}. 

\begin{figure}
\label{fig:pencil_beam}
\includegraphics[width=3.1in]{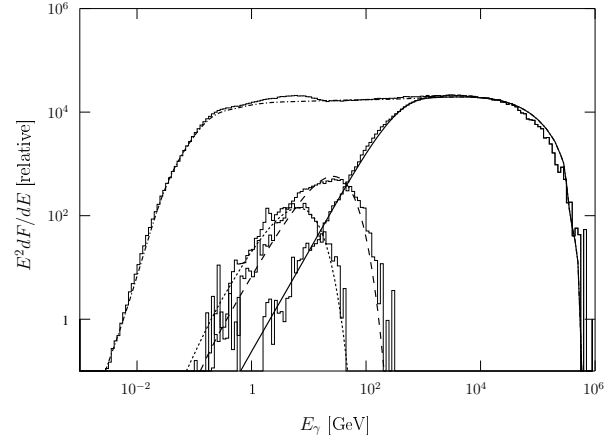}
\caption{Gamma-ray spectra from a pencil beam of protons, power law with index 2.0, observed at three different angles: head on (solid), $0.5^{\circ}$ (dashed) and $2^{\circ}$ (dotted). The thin line is for integration over the entire phase space. Histograms are the corresponding spectra from Monte Carlo simulations. The absolute normalization is relative to the density and distribution of target protons. Fluctuations in the histograms are due to low event statistics.}
\end{figure}

The gamma-ray spectrum due to a beam of protons along the z-axis with no spatial extension in the x-y plane, i.e.\ a pencil beam, has been calculated using the parameterization of $p_{t}$ distributions. The beam is viewed from three different angles, head on, $0.5^{\circ}$ and $2^{\circ}$, giving three different spectra. These is shown in Figure \ref{fig:pencil_beam} together with the total inclusive spectrum integrsted over the entire phase space. Each spectrum is integrated over the annular portion $(\theta,\theta + \Delta\theta)$ with $\Delta\theta=2'$. The normalization is relative to the density and distribution of target material. The histogram fluctuations are due to low event statistics.

\section{Starburst Galaxies}
A starburst galaxy is a galaxy with regions on highly enhanced star formation. Compared with a normal galaxy, the star-formation rate can be orders of magnitude higher in a starburst region. This also implies a higher rate of supernovae explosions and if the assumption that supernovae remnants accelerate the bulk of the cosmic rays in a galaxy is correct, then the total cosmic-ray flux in a starburst region is expected to be much higher than what is observed in the Milky Way. A starburst region also has a denser interstellar medium. This make starbursts the more likely candidates for detection of a non-active galaxy, even in TeV gamma rays.

Several predictions of expected gamma-ray fluxes from starburst galaxies have been made. \citet{article:Paglione_etal:1996} modeled NGC 253 and matched their results to data obtained with EGRET. The EGRET data gives a $2\sigma$ upper limit on the flux above 100 MeV of $8\times10^{-8}$ photons cm$^{-2}$ s$^{-1}$. Based on their calculations of the gamma-ray spectrum (see Figure 6 in \citep{article:Paglione_etal:1996}), the Fermi Large Area Telescope (Fermi-LAT) should be able to detect NGC 253 based on a one-year observation. Current generation ground based telescope will most likely have difficulties detecting this source. A detection was claimed by CANGAROO \citep{article:Itoh_etal:2002} but the spectrum is uncertain and H.E.S.S.\ collaboration has not confirmed the detection.

Perhaps an even better detection candidate is M82 in the northern hemisphere. The integrated flux above 100 GeV was estimated to be about $2\times 10^{-12}$ photons cm$^{-2}$ s$^{-1}$ \citep{article:Persic_etal:2008}. Based on this model, it is a possible candidate for detection by the ground-based telescopes MAGIC and VERITAS and it is a good candidate for the MAGIC II telescope. They also calculated the integrated flux above 100 MeV to be about $\times 10^{-8}$ photons cm$^{-2}$ s$^{-1}$ which is detectable by the Fermi-LAT over a one-year observation.

The author has noted that most calculations of the pion component of the gamma-ray spectrum is based on the outdated $p$-$p$ interaction models discussed earlier in this paper. In particular, they assume a constant inelastic $p$-$p$ cross section. As was shown by \citet{article:KamaeAbeKoi:2005}, inclusion of a logarithmically rising cross section significantly increases the gamma-ray yield at TeV energies.

\section{Conclusions and Future Work}
The $p$-$p$ interaction model presented here facilitate computation of gamma-ray spectra in both cases of both isotropic and anisotropic proton distributions. The formulae incorporate all important known features of the inelastic $p$-$p$ interaction up to about 500 TeV. Its inclusion in starburst models is expected to increase the gamma-ray yield by at least 25\%.

The author aims to further refine the $p$-$p$ interaction model and apply it to detailed modeling and analysis of several starburst galaxies, including prominent detection candidates M82 and NGC 253. A more rigorous study of several other starbursts will be carried out.


\begin{theacknowledgments}
The author acknowledges the contributions and guidance given by Professor Tuneyoshi Kamae while working at the Stanford Linear Accelerator Center and his thesis advisor Professor Per Carlson at the Royal Institute of Technology, Sweden. This work was in part supported by the U.S. Department of Energy under Grant DE-AC02-76SF00515, the Swedish GLAST consortium, and the Brinson Foundation.
\end{theacknowledgments}

\bibliographystyle{aipproc}   
\bibliography{refs}

\end{document}